\documentclass[12pt,a4paper]{article}
\pdfoutput=1
\usepackage{jcappub}
\usepackage{amsmath}
\usepackage{latexsym}
\usepackage{amsfonts}
\usepackage{graphicx}
\usepackage{mathrsfs}
\usepackage{epstopdf}

\title{Constant-roll tachyon inflation and observational constraints}

\author[a]{Qing Gao}
\author[b,1]{and Yungui Gong,\note{Corresponding author}}
\author[b]{Qin Fei}

\affiliation[a]{School of Physical Science and Technology, Southwest University, Chongqing 400715, China}

\affiliation[b]{School of Physics, Huazhong University of Science and Technology,
Wuhan, Hubei 430074, China}

\emailAdd{gaoqing1024@swu.edu.cn}
\emailAdd{yggong@hust.edu.cn}
\emailAdd{feiqin@hust.edu.cn}

\abstract{For the constant-roll tachyon inflation,
we derive the analytical expressions for the scalar and tensor power spectra, the scalar and tensor spectral tilts
and the tensor to scalar ratio to the first order of $\epsilon_1$ by using the method of Bessel function approximation. The derived $n_s$-$r$ results are compared with the observations,
we find that only the constant-roll inflation with $\eta_H$ being a constant is consistent with the observations
and observations constrain the constant-roll inflation to be slow-roll inflation.
The tachyon potential is also reconstructed for the constant-roll inflation which is consistent with the observations.
}

\arxivnumber{1801.09208}

\begin{document}

\maketitle
\section{Introduction}

The temperature and polarization measurements on the cosmic microwave background anisotropy
gave the constraints $n_s=0.9645\pm 0.0049$ (68\% C.L.)
and $r_{0.002}<0.10$ (95\% C.L.) \cite{Ade:2015lrj}.
If we take the number of $e$-folds before the end of inflation when a pivotal scale such as  $k_*=0.002$ Mpc$^{-1}$
crosses out the horizon, $N=60$, the observational results suggest that $n_s=1-2/N$.
This attractor behavior can be realized in chaotic inflation with the quadratic potential \cite{Linde:1983gd},
the T model with the potential $V(\phi)\sim \tanh^{2n}(\phi/\sqrt{6})$ \cite{Kallosh:2013hoa},
the E model with the potential $V(\phi)\sim \tanh^{2n}(\phi/\sqrt{6})$ \cite{Kallosh:2013maa},
the Starobinsky $R+R^2$ model \cite{Starobinsky:1980te},
Higgs inflation with the nonminimal coupling $\xi\phi^2 R$ in the strong coupling limit $\xi\gg 1$  \cite{Kaiser:1994vs,Bezrukov:2007ep},
and a class of inflationary models with nonminimal coupling to gravity \cite{Kallosh:2013tua,Yi:2016jqr,Nojiri:2017ncd}.
The attractor behavior also motivates the parametrization of $n_s$ and $r$ by $N$ and the reconstruction
of the inflationary potential with the parametrization by neglecting higher order corrections
\cite{Huang:2007qz,Mukhanov:2013tua,Roest:2013fha,Garcia-Bellido:2014gna,Garcia-Bellido:2014wfa,
Garcia-Bellido:2014eva,Creminelli:2014nqa,Boubekeur:2014xva,Binetruy:2014zya,
Barranco:2014ira,Galante:2014ifa,Gobbetti:2015cya,Chiba:2015zpa,Cicciarella:2016dnv,Lin:2015fqa,Nojiri:2010wj,
Odintsov:2016vzz,Choudhury:2017cos,Gao:2017uja,Jinno:2017jxc,Gong:2014cqa}.

If the potential of the inflaton is very flat,
then the inflaton almost stops rolling and the ultra slow-roll inflation is reached \cite{Tsamis:2003px,Kinney:2005vj}.
In the ultra slow-roll inflation, a large curvature perturbation at small scales may be generated to seed
primordial black holes \cite{Germani:2017bcs,Gong:2017qlj}.
More generally, the constant-roll inflation which includes the slow-roll inflation
with small rate of roll and the ultra slow-roll inflation was proposed \cite{Martin:2012pe,Motohashi:2014ppa}.
In constant-roll inflation, there exists exact solutions, the curvature perturbation may evolve on super-horizon
scales and the non-Gaussianity consistency relation may be violated, so the constant-roll inflation has richer physics
than the slow-roll inflation does.
For the constant-roll or ultra slow-roll inflation,
the slow-roll condition is violated, the curvature perturbation may not remain to
be a constant outside the horizon \cite{Jain:2007au,Namjoo:2012aa, Martin:2012pe,Motohashi:2014ppa},
so the slow-roll results cannot be applied \cite{Kinney:2005vj,Chongchitnan:2006wx,Martin:2012pe,Motohashi:2014ppa,Yi:2017mxs}.
For more discussion on constant-roll inflation and its reconstruction, please see
\cite{Motohashi:2017vdc,Motohashi:2017aob,Oikonomou:2017bjx,Odintsov:2017qpp,Nojiri:2017qvx,
Gao:2017owg,Dimopoulos:2017ged,Ito:2017bnn,Karam:2017rpw,Cicciarella:2017nls,Anguelova:2017djf}.

Apart from a canonical scalar field to drive inflation, an effective scalar field with nonlinear kinetic term
which describes the tachyon condensate in the string theory  \cite{Sen:2002nu,Sen:2002in}
also drives inflation and  gives the almost scale invariant power spectrum
\cite{Mazumdar:2001mm,Gibbons:2002md,Padmanabhan:2002cp,Frolov:2002rr,Garriga:1999vw,Hwang:2002fp,Steer:2003yu,Choudhury:2015hvr}.
The reconstruction of the tachyon potential with the help of the parametrization of $n_s$ and $r$ in terms of $N$
was discussed in \cite{Barbosa-Cendejas:2015rba,Fei:2017fub,Barbosa-Cendejas:2017pbo}.
The rolling tachyon on unstable D-branes in bosonic and superstring theories may behave as dark matter at late time \cite{Sen:2002in},
so it naturally leads the transition from early time inflation to late time matter domination.
Since the current observation is still
unable to address the nature of scalar field, it is interesting to study tachyon inflation and its physical implications.
In previous studies, tachyon inflation was considered under the slow-roll condition.
The slow-roll condition is violated in the constant-roll inflation when the constant rate of roll is not small,
the power spectra for both the scalar and tensor perturbations
derived under the slow-roll approximation cannot be applied to the constant-roll inflation.
In this paper, we discuss the constant-roll tachyon inflation and derive
the analytical formulae for the power spectra \footnote{While this work is in progress, the paper \cite{Mohammadi:2018oku} appeared, discussing
the power spectrum for one constant-roll inflationary model.}.
We then use the observational data to constrain the constant-roll tachyon inflationary models.
The reconstruction of the tachyon potential is also discussed.

The paper is organized as follows. In section \ref{sec2},
we first review the slow-roll tachyon inflation and introduce four different definitions of slow-roll parameters.
The scalar and tensor perturbations for constant-roll tachyon inflation are then derived.
In section \ref{sec3}, we derive the formulae for the scalar spectral tilt $n_s$ and
the tensor-to-scalar ratio $r$ for the four constant-roll inflationary models,
and use the observational data to constrain the models.
The reconstruction of the tachyon potential for the model with constant $\eta_H$ is presented in section \ref{sec4}.
The conclusions are drawn in section \ref{sec5}.

\section{Tachyon inflation}
\label{sec2}

We start with the effective action for the rolling tachyon
\begin{equation}
\label{tachyon1}
S_T=-\int d^4x \sqrt{-g}\,V(T)\sqrt{1+g^{\mu\nu}\partial_\mu T\partial_\nu T}.
\end{equation}
The string motivated potential $V(T)$ has a global maximum at $T=T_0$ and a minimum $V\rightarrow 0$ as $T\rightarrow \infty$.
Applying the Friedmann-Robertson-Walker metric for the homogeneous and isotropic spacetime,
we get the background equations of motion
\begin{equation}
\label{frweq1}
H^2=\frac{1}{3}\frac{V}{\sqrt{1-\dot T^2}},
\end{equation}
\begin{equation}
\label{frweq2}
\frac{\ddot T}{1-\dot T^2}+3H\dot T+\frac{V_{,T}}{V}=0.
\end{equation}
where $V_{,T}=dV/dT$, and we set $M_{pl}=1/\sqrt{8\pi G}=1$.
Combining eqs. \eqref{frweq1} and \eqref{frweq2}, we get
\begin{equation}
\label{acceq1}
\dot H=-\frac{3}{2}H^2\dot T^2.
\end{equation}
From eq. \eqref{acceq1}, we obtain the acceleration
\begin{equation}
\label{acceq2}
\frac{\ddot a}{a}=\dot H+H^2=H^2\left(1-\frac{3}{2}\dot T^2\right).
\end{equation}
So the occurrence of inflation $\ddot a>0$ is equivalent to $\dot T^2<2/3$.

\subsection{Slow-roll inflation}

Under the slow-roll approximations,
\begin{gather}
\label{slreq1}
\dot T^2\ll 1,\\
\label{slreq2}
\quad |\ddot T|\ll 3H|\dot T|,
\end{gather}
the background equations \eqref{frweq1} and \eqref{frweq2} for the tachyon become
\begin{gather}
\label{frweq3}
H^2\approx \frac{V}{3}, \\
\label{frweq4}
3H \dot T\approx -V_{,T}/V.
\end{gather}

\subsection{Slow-roll parameters}
In this subsection, we introduce several different definitions of the slow-roll parameters.
First, we introduce the horizon flow slow-roll parameters \cite{Schwarz:2001vv}
\begin{gather}
\label{slreq3}
\epsilon_0=\frac{H_o}{H},\\
\label{slreq4}
\epsilon_{i+1}=-\frac{d \ln|\epsilon_i|}{dN},
\end{gather}
where $H_o$ is an arbitrary constant. For the tachyon field,
the first two slow-roll parameters are \cite{Steer:2003yu}
\begin{gather}
\label{slreq5}
\epsilon_1=-\frac{\dot H}{H^2}=\frac{3}{2}\dot T^2,\\
\label{slreq6}
\epsilon_2=-\frac{d\ln\epsilon_1}{dN}=2\frac{\ddot T}{H\dot T}.
\end{gather}
By using these slow-roll parameters, the slow-roll conditions \eqref{slreq1} and \eqref{slreq2} are expressed as
$\epsilon_1\ll 1$ and $|\epsilon_2|\ll 1$, and inflation ends when $\epsilon_1=1$. Under the slow-roll approximations, we also get
\begin{gather}
\label{slreq7}
\epsilon_1\approx \frac{1}{2}\frac{V^2_{,T}}{V^3},\\
\label{slreq8}
\epsilon_2\approx -2\frac{V_{,TT}}{V^2}+3\frac{V^2_{,T}}{V^3}.
\end{gather}
The remaining number of $e$-folds $N(t)=\ln (a_f/a)$ before the end of inflation is
\begin{equation}
\label{neeq2}
N(t)=\int_t^{t_f} H(t)dt=\pm \sqrt{\frac{3}{2}}\, \int_T^{T_f}\frac{H}{\sqrt{\epsilon_1}}dT\approx \int^T_{T_f} \frac{V^2}{V_{,T}} dT,
\end{equation}
where the subscript $f$ denotes the end of inflation,
and the $\pm$ sign is the same as the sign of $\dot T$.
The last approximation is only valid under the slow-roll conditions.

Next, we introduce the Hubble flow slow-roll parameters \cite{Liddle:1994dx}
\begin{equation}
\label{hfslr1}
^n\beta_H=\frac{2}{3H^2}\left(\frac{(H_{,T})^{n-1}H^{(n+1)}}{H^n}\right)^{1/n},
\end{equation}
where $H^{(n)}=d^nH/dT^n$ and extra $1/H^2$ factor is added for the tachyon field. In terms of the
Hubble flow slow-roll parameters, the two first order slow-roll parameters are
\begin{equation}
\label{hfslr2}
\epsilon_H=\frac{2}{3H^2}\left(\frac{H_{,T}}{H}\right)^2=\epsilon_1,
\end{equation}
\begin{equation}
\label{hfslr3}
\eta_H=\frac{2 H_{,TT}}{3H^3}=2\epsilon_1-\frac{1}{2}\epsilon_2.
\end{equation}
In terms of the slow-roll parameter $\eta_H$, the slow-roll condition \eqref{slreq2} becomes $|\eta_H|\ll 1$. Under the slow-roll condition,
\begin{equation}
\label{hfslr21}
\eta_H\approx -\frac{1}{2}\frac{V_{,T}^2}{V^3}+\frac{V_{,TT}}{V^2}.
\end{equation}

In analogy with the canonical scalar field, we can also use $\ddot H$ to define the slow-roll parameter
\begin{equation}
\label{hddotslr2}
\epsilon_{2H}=-\frac{\ddot H}{2H\dot H}=\epsilon_1-\frac{1}{2}\epsilon_2.
\end{equation}
In terms of the slow-roll parameter $\epsilon_{2H}$, the slow-roll condition \eqref{slreq2} becomes $|\epsilon_{2H}|\ll 1$. Under the slow-roll condition,
\begin{equation}
\label{hfslr21}
\epsilon_{2H}\approx -\frac{V_{,T}^2}{V^3}+\frac{V_{,TT}}{V^2}.
\end{equation}

Finally, we introduce the slow-roll parameter
\begin{equation}
\label{tddotslr2}
\epsilon_{2T}=-\frac{\ddot T}{H\dot{T}(1-\dot{T}^2)}=-\frac{\epsilon_2}{2(1-2\epsilon_1/3)}.
\end{equation}
In terms of the slow-roll parameter $\epsilon_{2T}$, the slow-roll condition \eqref{slreq2} becomes $|\epsilon_{2T}|\ll 1$.
For a very flat potential with $V_{,T}\approx 0$, we get the ultra slow-roll inflation and $\epsilon_{2T}\approx 3$,
so this slow-roll parameter is useful for the discussion of the ultra slow-roll inflation.
Note that when slow-roll condition is satisfied, all the slow-roll parameters introduced above are small.

\subsection{The scalar perturbation}
In the flat gauge $\delta T(x,t)=0$, the gravitational action $\int d^4x \sqrt{-g}R$
plus the action \eqref{tachyon1} for the curvature perturbation $\delta g_{ij}=a^2(1+2\zeta)\delta_{ij}$ becomes
\begin{equation}
\label{tachyon3}
S=-\frac{3}{2}\int d^4x \left[a\dot T^2(\partial_i\zeta)^2-a^3\frac{\dot T^2}{1-\dot T^2}\dot\zeta^2\right].
\end{equation}
Using the canonically normalized field $v = z \zeta$,
the action (\ref{tachyon3}) becomes
\begin{equation}
\label{tachyon4}
S = \int d^3x d\tau \frac{1}{2} \left[ v{'^2} - c_s^2 (\partial_i v)^2 + \frac{z''}{z} v^2 \right],
\end{equation}
where the prime denotes the derivative with respect to the conformal time $\tau=\int dt/a$,
the effective sound speed is $c_s^2 = 1-\dot T^2=1-2\epsilon_1/3$ \cite{Garriga:1999vw}, and
\begin{equation}
\label{normvars}
z = \frac{\sqrt 3a\dot T}{\sqrt{1-\dot T^2}}.
\end{equation}

Now we introduce the quantum operator
\begin{equation}
\label{veq21}
\hat v(\tau ,\vec{x}) = \int \frac{d^3k}{( 2\pi)^3}\left[v_k(\tau)\hat a_k e^{i\vec{k} \cdot \vec{x}}+v_k^*(\tau)\hat a_k^\dag e^{-i\vec{k} \cdot \vec{x}}\right],
\end{equation}
with the creation and annihilation operators satisfying the standard commutation relations
\begin{equation}
\label{veq22}
\begin{split}
\left[\hat a_k,\ \hat a_{k'}^\dag\right]=(2\pi)^3\delta^3(\vec{k}-\vec{k'}),\\
\left[\hat a_k,\ \hat a_{k'}\right]=\left[\hat a_k^\dag,\ \hat a_{k'}^\dag\right]=0,
\end{split}
\end{equation}
By choosing the Bunch-Davies vacuum $\hat a_k|0\rangle=0$ \cite{bunchd}, the mode function $v_k$ obeys the normalization condition
\begin{equation}
\label{eq23}
v_k' v_k^* - v_k {v_k^*}' =  - i.
\end{equation}

Varying the action \eqref{tachyon4},
we obtain the Mukhanov-Sasaki equation for the mode function $v_k(\tau)$ \cite{Steer:2003yu},
\begin{equation}
\label{eq21}
v_k'' + \left( c_s^2 k^2 - \frac{z''}{z} \right)v_k = 0.
\end{equation}

To solve the Mukhanov-Sasaki equation \eqref{eq21}, we need the expression for $z''/z$.
In terms of the slow-roll parameters,
from the definition \eqref{normvars} we get \cite{Steer:2003yu}
\begin{equation}
\label{dotz1}
\dot z = Hz\left[1+\frac{\epsilon_2}{2\left(1-\frac{2}{3}\epsilon_1\right)}\right],
\end{equation}
\begin{equation}
\label{ddotz}
\frac{\ddot z}{z} = -H^2 \epsilon_1\left[1+\frac{\epsilon_2}{2\left(1-\frac{2}{3}\epsilon_1\right)}\right]+H^2 \left[1+\frac{\epsilon_2}{2\left(1-\frac{2}{3}\epsilon_1\right)}\right]^2
+\frac{H^2\epsilon_2^2\epsilon_1}{3\left(1-\frac{2}{3}\epsilon_1\right)^2}
+\frac{H\dot{\epsilon_2}}{2\left(1-\frac{2}{3}\epsilon_1\right)},
\end{equation}

\begin{equation}
\label{slreq7a}
z'=\frac{dz}{d\tau}=a\dot z=aHz\left[1+\frac{\epsilon_2}{2\left(1-\frac{2}{3}\epsilon_1\right)}\right],
\end{equation}
\begin{equation}
\label{zpp}
\frac{z''}{z}=a^2H^2\left[1+\frac{\epsilon_2}{2\left(1-\frac{2}{3}\epsilon_1\right)}\right]+\frac{a^2\ddot z}{z}.
\end{equation}
From the relation
\begin{equation}
\label{ah0}
\frac{d}{d\tau}\left(\frac{1}{aH}\right)=-1+\epsilon_1,
\end{equation}
we get
\begin{equation}
\label{ah}
\frac{1}{aH}=-\tau+\int \epsilon_1 d\tau=\tau(\epsilon_1-1)-\int\tau\frac{\epsilon_1}{d\tau}d\tau,
\end{equation}
Since
\begin{equation}
\label{dotep}
\dot{\epsilon_1}=H\epsilon_1\epsilon_2,
\end{equation}
so
\begin{equation}
\label{eps1res1}
\int \tau \frac{\epsilon_1}{d\tau}d\tau=\int aH\tau \epsilon_1\epsilon_2 d\tau.
\end{equation}
If $\epsilon_2$ is a constant,
to the first order of $\epsilon_1$, we get
\begin{equation}
\label{eps1res2}
\int \tau \frac{\epsilon_1}{d\tau}d\tau=-\epsilon_2\int \epsilon_1 d\tau.
\end{equation}
Combining eqs. \eqref{ah} and \eqref{eps1res2}, to the first order of $\epsilon_1$, for constant $\epsilon_2$ we obtain \cite{Yi:2017mxs}
\begin{equation}
\label{aheq1}
\frac{1}{aH}\approx \left(\frac{\epsilon_1}{1-\epsilon_2}-1\right)\tau.
\end{equation}
Because we derive the above result with the relation \eqref{dotep},
so the result \eqref{aheq1} does not apply to the case with $\epsilon_1$ being a constant or large.
Since we need the relation \eqref{aheq1}, we assume $\epsilon_2$ is a constant and $\epsilon_1$ is small
in this section for the convenience of discussion.
Substituting eq. \eqref{aheq1} into \eqref{zpp}, we can express $z''/z$ in terms of
a function of the slow-roll parameters $\epsilon_1$ and $\epsilon_2$ divided by $\tau^2$,
and we rewrite eq. \eqref{eq21} as
\begin{equation}
\label{eq21a}
v_k'' + \left( c_s^2 k^2 - \frac{\nu^2-1/4}{\tau^2} \right)v_k = 0,
\end{equation}
where
\begin{equation}
\label{slreq9}
\nu^2=\frac{1}{4}+\frac{z''}{z}\tau^2
\end{equation}
depends on the slow-roll parameters $\epsilon_1$ and $\epsilon_2$ only. For the slow-roll inflation,
$\epsilon_1$ and $\epsilon_2$ changes slowly, $\nu$ can be approximated as a constant. For the
constant-roll inflation, $\nu$ can also be approximated as a constant.
For either case, $\nu$ is almost a constant,
the solution to eq. \eqref{eq21a} for the mode function $v_k$ is the Hankel function of order $\nu$.
If $\epsilon_2$ is too large, then from eq. \eqref{dotep}, we see that $\dot{\epsilon}_1$ may not small,
and the Bessel function approximation may break down \cite{Yi:2017mxs}.
Here we don't consider this issue and leave it for future discussion.
By matching the Hankel function to the asymptotic plane wave solution with $k\rightarrow \infty$
which is consistent with the normalization condition \eqref{eq23}, we get the curvature perturbation
on super-horizon scales,
\begin{equation}
\label{vksol8}
|\zeta_k|=\frac{|v_k|}{z}=2^{\nu-2}\frac{\Gamma(\nu)}{\Gamma(3/2)}
\left[\frac{1}{aH}\left(1+\frac{\epsilon_1}{1-\epsilon_2}\right)\right]^{\frac{1}{2}-\nu}(c_s k)^{-\nu}/z.
\end{equation}
On super-horizon scales, $k\rightarrow 0$, the curvature perturbation $\zeta$ may not remain to be a constant.
In this paper, we focus on the usual situation that the curvature perturbation remains constant.
Therefore, the power spectrum of the scalar perturbation is
\begin{equation}
\label{pkeq1}
P_{\zeta}=\frac{k^3}{2\pi^2}|\zeta_k|^2=\left.\frac{2^{2\nu-3}}{2c_s\epsilon_1}\left[\frac{\Gamma(\nu)}{\Gamma(3/2)}\right]^2
\left(1+\frac{\epsilon_1}{1-\epsilon_2}\right)^{1-2\nu}\left(\frac{H}{2\pi}\right)^2\left(\frac{c_s k}{aH}\right)^{3-2\nu}\right|_{c_sk=aH}.
\end{equation}
The amplitude of the scalar perturbation is
\begin{equation}
\label{aseq1}
A_s=\left.\frac{2^{2\nu-3}}{c_s}\left[\frac{\Gamma(\nu)}{\Gamma(3/2)}\right]^2
\left(\frac{1-\epsilon_2+\epsilon_1}{1-\epsilon_2}\right)^{1-2\nu}\frac{H^2}{8\pi^2\epsilon_1}\right|_{c_sk=aH}.
\end{equation}
The scalar spectral tilt is
\begin{equation}
\label{nseq1}
n_s-1= \frac{d\ln P_\zeta }{d\ln k} =3-2\nu.
\end{equation}

\subsection{The tensor perturbation}
For the tensor perturbation $\delta g_{ij}=a^2\gamma_{ij}$, to the second order,
the gravitational action plus the action \eqref{tachyon1} becomes
\begin{equation}
\label{tachyon7}
S=\frac{1}{8}\int d^4x\left[a^3(\dot\gamma_{ij})^2-a(\gamma_{ij,k})^2\right],
\end{equation}
where $\gamma_{ij}=\sum_{s=+,\times}e_{ij}^s \gamma^s$.
Following the same procedure as that in the scalar perturbation,
we introduce the normalized field $u=a\gamma/\sqrt{2}$, and get eq. \eqref{eq21a} with $v$ replaced by $u$,
and $\nu$ replaced by $\mu$, where
\begin{equation}
\label{mueq11}
\mu^2=\frac{1}{4}+\frac{a''}{a}\tau^2,
\end{equation}
and
\begin{equation}
\label{mueq12}
\frac{a''}{a}=a^2H^2(2-\epsilon_1),
\end{equation}
so the tensor spectrum is
\begin{equation}
\label{pteq1}
P_T=2^{2\mu}\left[\frac{\Gamma(\mu)}{\Gamma(3/2)}\right]^2
\left(1+\frac{\epsilon_1}{1-\epsilon_2}\right)^{1-2\mu}\left(\frac{H}{2\pi}\right)^2\left(\frac{ k}{aH}\right)^{3-2\mu}.
\end{equation}
The tensor spectral tilt is
\begin{equation}
\label{nteq1}
n_T=\frac{d\ln P_T}{d\ln k}=3-2\mu.
\end{equation}
Combining eqs. \eqref{aseq1} and \eqref{pteq1}, to the first order of $\epsilon_1$, we get the tensor to scalar ratio
\begin{equation}
\label{req1}
r=2^{2(\mu-\nu)+4}\left[\frac{\Gamma(\mu)}{\Gamma(\nu)}\right]^2 \epsilon_1.
\end{equation}

\section{The constant-roll inflationary models}
\label{sec3}

\subsection{Constant $\epsilon_2$}

In this subsection, we consider the case that $\epsilon_2$ is a constant and derive the formulae for $n_s$ and $r$
to the first order of $\epsilon_1$.
From eq. \eqref{aheq1}, to the first order of $\epsilon_1$, we get
\begin{equation}
\label{ah1}
aH\approx-\frac{1}{\tau}\left(1+\frac{\epsilon_1}{1-\epsilon_2}\right).
\end{equation}
Using $\dot\epsilon_2=0$ and combining eqs. \eqref{ddotz}, \eqref{zpp}, \eqref{slreq9} and \eqref{ah1},
to the first order of $\epsilon_1$, we obtain
\begin{equation}
\label{nu1}
\nu\approx\frac{1}{2}|3+\epsilon_2|+\frac{(4\epsilon_2^3-4\epsilon_2^2-27\epsilon_2-18)\epsilon_1}
{6|3+\epsilon_2|(\epsilon_2-1)},
\end{equation}
\begin{equation}
\label{mu1}
\mu\approx \frac{3}{2}+\frac{3+\epsilon_2}{3(1-\epsilon_2)}\epsilon_1.
\end{equation}
Substituting eqs. \eqref{nu1} and \eqref{mu1} into eqs. \eqref{nseq1} and \eqref{req1},
to the first order of $\epsilon_1$, we derive that
\begin{equation}
\label{nsa}
n_s\approx4-|3+\epsilon_2|+\frac{(18-4\epsilon_2^3+4\epsilon_2^2+27\epsilon_2)\epsilon_1}
{3|3+\epsilon_2|(\epsilon_2-1)},
\end{equation}
\begin{equation}
\label{ra}
r\approx2^{3-|3+\epsilon_2|}\left(\frac{\Gamma[3/2]}{\Gamma[|3+\epsilon_2|/2]}\right)^2 \, 16\epsilon_1.
\end{equation}
If the slow-roll condition is satisfied, $|\epsilon_2|\ll 1$, the results become $n_s=1-2\epsilon_1-\epsilon_2$
and $r=16\epsilon_1$ \cite{Garriga:1999vw,Hwang:2002fp,Steer:2003yu,Fei:2017fub},
so the results for the slow-roll tachyon inflation are recovered. To the first order of $\epsilon_1$,
the result \eqref{nsa} is different from that for the canonical scalar field found in \cite{Gao:2018cpp}.

\begin{figure}[htbp]
\centering
\includegraphics[width=0.6\textwidth]{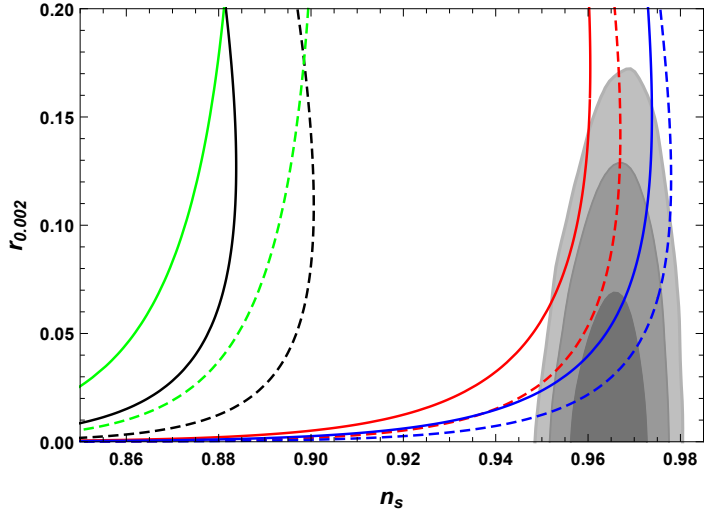}
\caption{The marginalized 68\%, 95\% and 99.8\% confidence level contours for
$n_s$ and $r$ from Planck 2015 data \citep{Ade:2015lrj} and the observational
constraints on $n_s-r$ for different constant-roll inflationary models.
The solid and dashed lines represent $N=50$ and $N=60$, respectively.
The green lines denote the model with constant $\epsilon_{2T}$,
the black lines denote the model with constant $\epsilon_2$,
the red lines denote the model with constant $\epsilon_{2H}$,
and the blue lines denote the model with constant $\eta_H$.}
\label{tachyon}
\end{figure}

Since $\epsilon_2$ is a constant, from the definition \eqref{slreq6}, we get
\begin{equation}
\label{epsna1}
\epsilon_1(N)=C\exp(-\epsilon_2N),
\end{equation}
where $C$ is an integration constant. At the end of inflation, $N=0$, $\epsilon_1(N)=1$, so $C=1$.
Substituting eq. \eqref{epsna1} into eqs. \eqref{nsa} and \eqref{ra}, we can calculate $n_s$ and $r$
for the model with constant $\epsilon_2$, and
the results along with the Planck 2015 constraints \citep{Ade:2015lrj} are shown in figure \ref{tachyon}.
In figure \ref{tachyon}, we plot the results by varying $\epsilon_2$ with $N=50$ and $N=60$,
and the black lines denote the results for the model with constant $\epsilon_2$. From figure \ref{tachyon},
we see that the model is ruled out by observations at the $3\sigma$ C.L.

\subsection{Constant $\epsilon_{2H}$}

In this subsection, we consider the case that $\epsilon_{2H}$ is a constant and derive the formulae for $n_s$ and $r$
to the first order of $\epsilon_1$.
From eq. \eqref{hddotslr2}, we have
\begin{equation}
\label{epsilon2h}
\epsilon_{2}=2(\epsilon_1-\epsilon_{2H}),
\end{equation}
\begin{equation}
\label{doteps2h}
\dot\epsilon_2=4H\epsilon_1(\epsilon_1-\epsilon_{2H}).
\end{equation}
Replacing $\epsilon_2$ with $\epsilon_{2H}$ by the relation \eqref{epsilon2h} and using the result \eqref{doteps2h} for $\dot\epsilon_2$,
to the first order of $\epsilon_1$, we get
\begin{equation}
\label{ah2}
aH\approx-\frac{1}{\tau}\left(1+\frac{\epsilon_1}{1+2\epsilon_{2H}}\right),
\end{equation}
\begin{equation}
\label{nu2}
\nu\approx\frac{1}{2}|3-\epsilon_{2H}|+\frac{(16\epsilon_{2H}^3-16\epsilon_{2H}^2-21\epsilon_{2H}+18)\epsilon_1}
{3|3-\epsilon_{2H}|(2\epsilon_{2H}+1)},
\end{equation}
\begin{equation}
\label{mu2}
\mu\approx \frac{3}{2}+\frac{3-2\epsilon_{2H}}{3(1+2\epsilon_{2H})}\epsilon_1.
\end{equation}
Substituting eqs. \eqref{nu2} and \eqref{mu2} into eqs. \eqref{nseq1} and \eqref{req1}, to the first order of $\epsilon_1$, we obtain
\begin{equation}
\label{nsb}
n_s\approx4-|3-2\epsilon_{2H}|+\frac{(-32\epsilon_{2H}^3+32\epsilon_{2H}^2+42\epsilon_{2H}-36)\epsilon_1}
{3|3-2\epsilon_{2H}|(2\epsilon_{2H}+1)},
\end{equation}
\begin{equation}
\label{rb}
r\approx2^{3-|3-2\epsilon_{2H}|}\left(\frac{\Gamma[3/2]}{\Gamma[|3-2\epsilon_{2H}|/2]}\right)^2 16\epsilon_{1}.
\end{equation}
In the slow-roll limit, $|\epsilon_{2H}|\ll 1$, we get $n_s=1-4\epsilon_1+2\epsilon_{2H}$ and $r=16\epsilon_1$
which are consistent with slow-roll results.

Since $\epsilon_{2H}$ is a constant, from the definition \eqref{slreq6} and the condition $\epsilon_1(N=0)=1$, we derive that
\begin{equation}
\label{epsnb1}
\epsilon_1(N)=\frac{\epsilon_{2H}\exp(2\epsilon_{2H}N)}{\exp(2\epsilon_{2H}N)+\epsilon_{2H}-1},
\end{equation}
Substituting eq. \eqref{epsnb1} into eqs. \eqref{nsb} and \eqref{rb}, we can calculate $n_s$ and $r$
for the model with constant $\epsilon_{2H}$, and
the results along with the Planck 2015 constraints \citep{Ade:2015lrj} are shown in figure \ref{tachyon}.
In figure \ref{tachyon}, we plot the results by varying $\epsilon_{2H}$ with $N=50$ and $N=60$,
and the red lines denote the results for the model with constant $\epsilon_{2H}$. From figure \ref{tachyon},
we see that the model is inconsistent with the observations at the $1\sigma$ C.L.

\subsection{Constant $\eta_H$}
For the model with constant $\eta_H$, from eqs. \eqref{hfslr3} and \eqref{dotep} we get
\begin{equation}
\label{etah}
\epsilon_2=2(2\epsilon_1-\eta_H),
\end{equation}
\begin{equation}
\label{doteps2c}
\dot\epsilon_2=8H\epsilon_1(2\epsilon_1-\eta_H).
\end{equation}
Replacing $\epsilon_2$ with $\eta_{H}$ by the relation \eqref{etah} and using the result \eqref{doteps2c} for $\dot\epsilon_2$,
to the first order of $\epsilon_1$, we have
\begin{equation}
\label{ah3}
aH\approx-\frac{1}{\tau}\left(1+\frac{\epsilon_1}{1+2\eta_{H}}\right),
\end{equation}
\begin{equation}
\label{nu3}
\nu\approx\frac{1}{2}|3-2\eta_{H}|+\frac{(16\eta_{H}^3-40\eta_{H}^2-15\eta_{H}+27)\epsilon_1}
{3|3-2\eta_{H}|(2\eta_{H}+1)},
\end{equation}
\begin{equation}
\label{mu3}
\mu\approx \frac{3}{2}+\frac{3-2\eta_{H}}{3(1+2\eta_{H})}\epsilon_1.
\end{equation}
Substituting eqs. \eqref{nu3} and \eqref{mu3} into eqs. \eqref{nseq1} and \eqref{req1}, to the first order of $\epsilon_1$, we obtain
\begin{equation}
\label{nsc}
n_s\approx4-|3-2\eta_{H}|+\frac{(-32\eta_{H}^3+80\eta_{H}^2+30\eta_{H}-54)\epsilon_1}
{3|3-2\eta_{H}|(2\eta_{H}+1)},
\end{equation}
\begin{equation}
\label{rc}
r\approx2^{3-|3-2\eta_{H}|}\left(\frac{\Gamma[3/2]}{\Gamma[|3-2\eta_{H}|/2]}\right)^2\,16\epsilon_{1}.
\end{equation}
In the slow-roll limit, $|\eta_H|\ll 1$, we get $n_s=1-6\epsilon_1+2\eta_H$ and $r=16\epsilon_1$ which are consistent with the slow-roll results.
The slow-roll results
are the same as those for the canonical scalar field with $\eta_V$.

Since $\eta_{H}$ is a constant, from the definition \eqref{slreq6} and the condition $\epsilon_1(N=0)=1$, we get
\begin{equation}
\label{epsnc1}
\epsilon_1(N)=\frac{\eta_{H}\exp(2\eta_{H}N)}{2\exp(2\eta_{H}N)+\eta_{H}-2},
\end{equation}
Plugging eq. \eqref{epsnc1} into eqs. \eqref{nsc} and \eqref{rc}, we express $n_s$ and $r$ in terms of $N$ and $\eta_H$.
By choosing $N=50$ and $N=60$, and varying the value of $\eta_H$,
we plot the $n_s$-$r$ results for the model with constant $\eta_H$
along with the Planck 2015 constraints \citep{Ade:2015lrj} in figure \ref{tachyon}.
The blue lines denote the $n_s$-$r$ results for the model with constant $\eta_H$. From figure \ref{tachyon},
we see that the model with constant $\eta_H$ is consistent with the observations at $1\sigma$ C.L.
For $N=50$, the $1\sigma$ constraint is $-0.0135<\eta_H<-0.0036$,
the $2\sigma$ constraint is $-0.0184<\eta_H<0.006$,
and the $3\sigma$ constraint is $-0.0207<\eta_H<0.0146$.
For $N=60$, the $1\sigma$ constraint is $-0.018<\eta_H<-0.006$,
the $2\sigma$ constraint is $-0.0212<\eta_H<0.0013$,
and the $3\sigma$ constraint is $-0.023<\eta_H<0.007$.
If we take $\eta_H=-0.009$ and $N=60$, we get $\epsilon_1=0.0023$, $n_s=0.968$, $r=0.036$.
Since observations require that $\epsilon_1$ and $\eta_H$ are both small,
so the slow-roll condition is satisfied and this constant-roll inflation with constant $\eta_H$ is also a slow-roll inflation.
If we use the slow-roll formulae to fit the observations,
the $1\sigma$ constraint is $-0.014<\eta_H<-0.0039$, the $2\sigma$ constraint is $-0.018<\eta_H<0.0068$,
and the $3\sigma$ constraint is $-0.02<\eta_H<0.0168$ for $N=50$.
For $N=60$, the $1\sigma$ constraint is $-0.018<\eta_H<-0.0067$,
the $2\sigma$ constraint is $-0.021<\eta_H<0.0015$, and the $3\sigma$ constraint is $-0.023<\eta_H<0.01$.
The $2\sigma$ and $3\sigma$ upper bounds given by the slow-roll formulae are larger than those given by the
constant-roll formulae, so even in the slow-roll regime, the results are not exactly the same, but the constant-roll
formulae \eqref{nsc} and \eqref{rc} are more accurate.

\subsection{Constant $\epsilon_{2T}$}

For the model with constant $\epsilon_{2T}$, from eqs. \eqref{tddotslr2} and \eqref{dotep} we get
\begin{equation}
\label{epsilon2t}
\epsilon_2=-2\epsilon_{2T}\left(1-\frac{2}{3}\epsilon_1\right),
\end{equation}
\begin{equation}
\label{doteps2d}
\dot\epsilon_2=-\frac{8}{3}H\epsilon^2_{2T}\epsilon_1\left(1-\frac{2}{3}\epsilon_1\right).
\end{equation}
Replacing $\epsilon_2$ with $\epsilon_{2T}$ by the relation \eqref{epsilon2t} and using the result \eqref{doteps2d} for $\dot\epsilon_2$,
to the first order of $\epsilon_1$, we obtain
\begin{equation}
\label{ah4}
aH\approx-\frac{1}{\tau}\left(1+\frac{\epsilon_1}{1+2\epsilon_{2T}}\right),
\end{equation}
\begin{equation}
\label{nu4}
\nu\approx\frac{1}{2}|3-2\epsilon_{2T}|+\frac{(4\epsilon_{2T}^2-7\epsilon_{2T}+3)\epsilon_1}
{|3-2\epsilon_{2T}|(2\epsilon_{2T}+1)},
\end{equation}
\begin{equation}
\label{mu4}
\mu\approx \frac{3}{2}+\frac{3-2\epsilon_{2T}}{3(1+2\epsilon_{2T})}\epsilon_1.
\end{equation}
Plugging the results \eqref{nu4} and \eqref{mu4} into
eqs. \eqref{nseq1} and \eqref{req1}, we have
\begin{equation}
\label{nsd}
n_s\approx4-|3-2\epsilon_{2T}|+\frac{2(4\epsilon_{2T}^2-7\epsilon_{2T}+3)\epsilon_1}
{3|3-2\epsilon_{2T}|(2\epsilon_{2T}+1)},
\end{equation}
\begin{equation}
\label{rd}
r\approx2^{3-|3-2\epsilon_{2T}|}\left(\frac{\Gamma[3/2]}{\Gamma[|3-2\epsilon_{2T}|/2]}\right)^2 \, 16\epsilon_{1}.
\end{equation}
In the slow-roll limit, we get $|\epsilon_{2T}| \ll 1$, $n_s=1+2\epsilon_1/3+2\epsilon_{2T}$.

For constant $\epsilon_{2T}$, from the definition \eqref{slreq6} and the condition $\epsilon_1(N=0)=1$, we derive that
\begin{equation}
\label{epsnd1}
\epsilon_1(N)=\frac{3}{\exp(-2\epsilon_{2T}N)+2}.
\end{equation}
Substituting eq. \eqref{epsnd1} into eqs. \eqref{nsd} and \eqref{rd},
we express $n_s$ and $r$ in terms of $N$ and $\epsilon_{2T}$.
By choosing $N=50$ and $N=60$, and varying the value of $\epsilon_{2T}$,
we plot the $n_s$-$r$ results for the model with constant $\epsilon_{2T}$
along with the Planck 2015 constraints \citep{Ade:2015lrj} in figure \ref{tachyon}.
The green lines denote the results for the model with constant $\epsilon_{2T}$.
From figure \ref{tachyon}, we see that the model with constant $\epsilon_{2T}$ is
excluded by the observations at the $3\sigma$ C.L.

\section{The reconstruction of the potential}
\label{sec4}

\begin{figure}[htbp]
\centering
\includegraphics[width=0.6\textwidth]{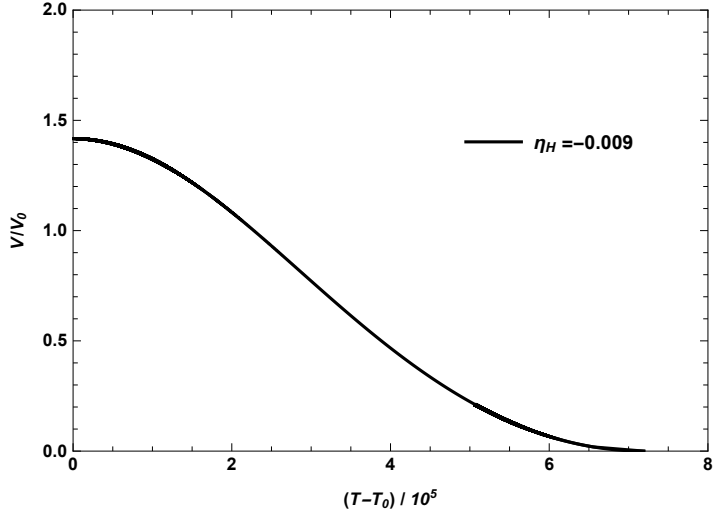}
\caption{The reconstructed potential normalized by $V_0=1.0386\times10^{-9}$.}
\label{VT}
\end{figure}

From the analysis in the previous section, we see that only the model with constant $\eta_H$ is
consistent with the observations and it is constrained to be a slow-roll inflation.
In this section, we follow the procedure presented in \cite{Fei:2017fub} for the slow-roll inflation to reconstruct the tachyon
potential with constant $\eta_H$. Combining eqs. \eqref{slreq7} and \eqref{neeq2}, we get
\begin{equation}
\label{eps1vrel1}
\epsilon_1\approx\frac{V_{,N}}{2V},
\end{equation}
where $V_{,N}=dV/dN$.
Substituting eq. \eqref{epsnc1} into \eqref{eps1vrel1}, we obtain
\begin{equation}
\label{conseq5}
V(N)\approx V_0\left|\eta_H+2\exp(2\eta_H N)-2\right|^{\frac{1}{2}},
\end{equation}
and
\begin{equation}
\label{conseq5a}
A_s\approx\frac{V}{24\pi^2\epsilon_1}=\frac{V_0|\eta_H+2\exp(2\eta_H N)-2|}{24\pi^2\epsilon_1}.
\end{equation}
If we take $\eta_H=-0.009$, $A_s=2.2\times10^{-9}$ and $N=60$, we get $V_0=1.0386\times10^{-9}$.
From the relation
\begin{equation}
\label{tneq1}
dT\approx \pm \frac{\sqrt{V_{,N}}}{V} dN,
\end{equation}
we get
\begin{equation}
\label{conseq7}
T-T_0\approx\sqrt{\frac{2}{|\eta_H|V_0}}\frac{\exp(\eta_H N)}{|\eta_H-2|^{3/4}}\, _2F_1\left(\frac{1}{2},\frac{3}{4};\frac{3}{2};\frac{2\exp(\eta_H N)}{2-\eta_H}\right).
\end{equation}
If we take $\eta_H=-0.009$, $A_s=2.2\times10^{-9}$ and $N=60$,
we find that the field excursion is $\Delta T=T_*-T_f=1.76\times 10^5$
and this result is consistent with the lower bound on the field excursion derived in \cite{Fei:2017fub}.
Combining eqs. \eqref{conseq5} and \eqref{conseq7},
we can obtain the potential $V(T)$ and the reconstructed potential for $\eta_H=-0.009$ is shown in figure \ref{VT}.
From figure \ref{VT}, we see that $V(T)$ has a maximum at $T=T_0$ and $V(T)\rightarrow 0$ as $T\rightarrow 0$,
this property is consistent with that of the string inspired potential.
Because it is difficult to obtain an analytical expression for $N$ in terms of $T$ from eq. \eqref{conseq7} in general,
here we give the analytical behavior of $V(T)$ around $T=T_0$. As $T\rightarrow T_0$, from eq. \eqref{conseq7}, we get
\begin{equation}
\label{conseq9}
e^{2\eta_H N}\approx \frac{1}{2}|2-\eta_H|^{3/2}|\eta_H|V_0(T-T_0)^2.
\end{equation}
Substituting eq. \eqref{conseq9} into eq. \eqref{conseq5}, we obtain the potential around $T=T_0$,
\begin{equation}
\label{conseq10}
V(T)\approx|2-\eta_H|^{1/2}V_0\left[1-\frac{1}{2}|2-\eta_H|^{1/2}|\eta_H|V_0(T-T_0)^2\right].
\end{equation}

\section{Conclusions}
\label{sec5}

We introduce four different definitions for the slow-roll parameters. For these four different constant-roll inflationary models,
we derive the analytical expressions for the scalar and tensor power spectra, the scalar and tensor spectral tilts
and the tensor to scalar ratio to the first order of $\epsilon_1$ by using the method of Bessel function approximation.
These results reduce to those for slow-roll inflation if slow-roll conditions are satisfied.
We also use the observational data to constrain the constant-roll inflationary models, and we find that
the constant-roll inflationary models with constant $\epsilon_2$ or constant $\epsilon_{2T}$
are ruled out by the observations at the $3\sigma$ C.L.
The model with constant $\epsilon_{2H}$ is inconsistent with the observations at the $1\sigma$ C.L.
The model with constant $\eta_H$ is consistent with the observations,
and the 1$\sigma$ constraint is $-0.0135<\eta_H<-0.0036$ if we take $N=50$;
the 1$\sigma$ constraint is $-0.018<\eta_H<-0.006$ if we take $N=60$.
Since the observational constraints tell us that $|\eta_H|\ll 1$, so the slow-roll conditions are satisfied
and the constant-roll inflation is also a slow-roll inflation.
Following the reconstruction procedure for the slow-roll inflation, we reconstruct the tachyon potential
for the model with constant $\eta_H$. The reconstructed tachyon potential satisfies the property for the string inspired potential,
but the possible origin of the potential from string theory needs further study.
If we choose $\eta_H=-0.009$, $A_s=2.2\times10^{-9}$ and $N=60$,
we get $\epsilon_1=0.0023$, $n_s=0.968$, $r=0.036$, $V_0=1.0386\times10^{-9}$ and $\Delta T=T_*-T_f=1.76\times 10^5$.
The field excursion for the tachyon is consistent with the general lower bound.
Although $\eta_H$ is constrained to be small and slow-roll inflation applies, the results for
constant-roll inflation are more general and have broad applications.

\begin{acknowledgments}

This research was supported in part by the National Natural Science
Foundation of China under Grant Nos. 11605061 and 11475065,
the Major Program of the National Natural Science Foundation of China under Grant No. 11690021, the Fundamental Research Funds for the Central Universities under Grant Nos. XDJK2017C059 and SWU116053.

\end{acknowledgments}


\providecommand{\href}[2]{#2}\begingroup\raggedright\endgroup

\end{document}